# Membrane buckling induced by curved filaments

Martin Lenz,[1,*] Daniel J. G. Crow,[1] and Jean-François Joanny[1]

[1]*Institut Curie, Centre de Recherche, Laboratoire Physico-Chimie Curie,
Paris, F-75248 France; CNRS, UMR 168, Paris,
F-75248 France; Université Pierre et Marie Curie-Paris6, UMR 168, Paris, F-75005 France*

(Dated: November 11, 2018)

We present a novel buckling instability relevant to membrane budding in eukaryotic cells. In this mechanism, curved filaments bind to a lipid bilayer without changing its intrinsic curvature. As more and more filaments adsorb, newly added ones are more and more strained, which destabilizes the flat membrane. We perform a linear stability analysis of filament-dressed membranes and find that the buckling threshold is within reasonable *in vivo* parameter values. We account for the formation of long tubes previously observed in cells and in purified systems. We study strongly deformed dressed membranes and their bifurcation diagram numerically. Our mechanism could be validated by a simple experiment.

Eukaryotic cells are highly compartmentalized, and many of their confining structures are made of lipid bilayers. In order to maintain the exchanges essential for their proper functioning, cells thus need tools that modulate the shape and topology of these membranes. Such tools may be proteins that self-assemble to form tubes in solution [1] and can impose this intrinsic tubular shape on membranes [2]. In some other physically interesting cases, the structure of the protein does not suggest an obvious tubulation mechanism, as for the Endosomal Sorting Complex Required for Transport III (ESCRT-III) [3]. This protein complex is implicated in the formation of multivesicular bodies [4], HIV budding [5] and cytokinesis [6], three processes which involve deformation of the membrane into a bud and/or severing off the resulting membrane protrusion from the inside. Deep-etch electron micrographs of COS-7 cells overexpressing hSnf-7, one of the constitutive proteins of ESCRT-III, reveal circular arrays of curved hSnf-7 polymers under the plasma membrane [Fig. 1(a)] [7]. This is evidence of the strong affinity of these filaments for the membrane [8] and for each other [9], as well as of their intrinsic curvature. When an ATP-hydrolysis deficient mutant of VPS4 —an ATPase involved in the disassembly of ESCRT-III filaments [9, 10]— is present, long membrane-covered tubes of hSnf-7 filaments are observed [Fig. 1(c)]. Similar structures appear in *in vitro* systems using purified proteins [11]. This suggests that tubes always form *in vivo*, but that in the presence of normal VPS4 alone they are immediately cut off the membrane to form vesicles. In this Letter we propose that this flat-to-tubular transition is a general feature of systems where curved filaments with attractive interactions bind to a membrane, and study this physical effect akin to the buckling of a rod (Fig. 1).

We consider an infinite, initially flat lipid bilayer parametrized by its radial coordinate $r$. A subdomain $r_i < r < r_e$ of this surface is bound to an array of filaments (Fig. 1). The dressed membrane is then put into contact with the cytoplasm, which acts as a reservoir of filaments. In the following, we consider only axisymmetric configurations [19] and assume that the dressed membrane is very thin. We write the free energy of the dressed membrane as:

$$\mathcal{F} = \int_{r \in \mathbb{R}^+} \left( \frac{\kappa}{2} c^2 + \sigma \right) d\mathcal{A} + 2\pi\gamma(r_i + r_e) \\ + \int_{r \in [r_i, r_e]} \left[ \frac{k}{2}\left(\frac{1}{r} - \frac{1}{r_0}\right)^2 - \mu \right] d\mathcal{A}. \quad (1)$$

The first term is the Helfrich free energy of the membrane, with bending modulus $\kappa$, local total curvature $c$ and tension $\sigma$ [12]. The second term represents the attractive interactions between filaments, characterized by a line tension $\gamma$. We assume that the filaments are closely packed; their surface density is thus constant throughout the array. The last term represents the free energy of the filaments. They have a preferred curvature $r_0^{-1}$, and due to the cylindrical symmetry their actual curva-ture is $r^{-1}$. A Taylor expansion about $r_0^{-1}$ to second order yields the filament stiffness $k$. We denote by $\mu$ the chemical potential difference between hSnf-7 in the cytoplasm and bound to the membrane. *In vivo*, the circular filaments pictured in Fig. 1 are not actually continuous and can be made of several consecutive shorter filaments. It is also possible that the hSnf-7 filaments are "living" polymers and exchange monomers with the cytoplasm. Therefore we consider that filaments of any length are always available, and that their chemical potential per monomer does not depend on their length, so that $\mu$ is uniform throughout the filament array. We ignore holes in the array resulting from thermal fluctuation, which is correct in the limit of large binding energies [20]. Eq. (1) also ignores the up-down asymmetry of the dressed membrane, a simplification discussed later. Finally, we define the scaled filament stiffness $K = k/\kappa$ and membrane tension $\Sigma = (\sigma - \mu)r_0^2/\kappa + k/(2\kappa)$.

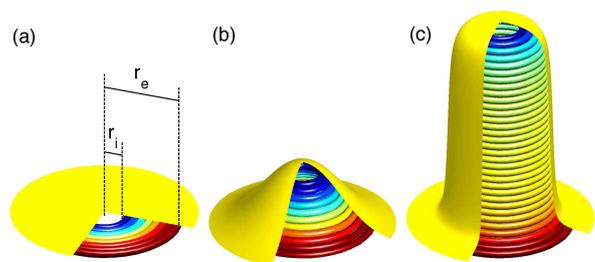

Figure 1: (color online) Illustration of the proposed buckling mechanism. Overbent filaments are represented in blue, underbent filaments in red and the membrane in yellow. Wedges of the membrane were removed for visualization. (a) Curved filaments with an affinity for each other and the membrane form membrane-bound circular arrays. The tension and bending modulus of the membrane tend to stabilize flat arrays. (b) Buckling, on the other hand, allows the binding of more filaments and the relaxation of those already bound to their preferred (yellow) radius. These stabilizing and destabilizing effects balance at the buckling threshold. (c) The formation of long tubes allows the binding of an arbitrarily large number of filaments close to their preferred radius.

We first consider the stability of flat arrays of filaments [Fig. 1(a)]. In Ref. [7], it is observed that these arrays have a finite, rather well-defined external radius $r_e$. We attribute this feature to a chemical equilibrium between hSnf-7 in the array and in solution. Minimizing $\mathcal{F}$ with respect to $r_e$ for a flat membrane ($c = 0$ and $d\mathcal{A} = 2\pi r\, dr$), one finds that the array has a finite external equilibrium radius only if $k/2r_0^2 > \mu$, i.e. only if it is more favorable for a filament to be in solution than bound to the rim of a very large ($r_e \to +\infty$) array. Line tension will shrink the array and make $r_e$ vanish unless

$$\gamma < \frac{k}{r_0}\left(1 - \sqrt{1 - \frac{2\mu r_0^2}{k}}\right). \qquad (2)$$

Under these assumptions we always have $0 < r_i < r_e$ and

$$\frac{r_e}{r_0} = \frac{k - r_0\gamma}{k - 2r_0^2\mu} + \sqrt{\left(\frac{k - r_0\gamma}{k - 2r_0^2\mu}\right)^2 - \frac{k}{k - 2r_0^2\mu}}. \qquad (3)$$

We now discuss the buckling of filament-dressed membranes [Fig. 1(b)]. Experimentally, it is observed that the typical length scale of a hSnf-7 protrusion is much larger than $r_i$ and smaller than $r_e$. We therefore assume for simplicity that $r_i = 0$ and $r_e \to +\infty$. We parametrize the dressed membrane by its altitude $z(r)$. The equilibrium states are the solutions of the force balance equation $\frac{\delta\mathcal{F}}{\delta z(r)} = 0$ with boundary conditions $\frac{dz}{dr}(0) = 0$ and $\frac{dz}{dr}(+\infty) = 0$. Therefore, $z(r)$ is defined up to an arbitrary additive constant. As in the case of a buckling rod [13], the buckling threshold is the set of parameters where non-zero solutions of the linearized force balance equation satisfying the boundary conditions exist. This equation reads:

$$z''' + \frac{z''}{R} - \left(\frac{1}{4} - \frac{1 + \sqrt{4+2K} + 2\nu}{2R} + \frac{2+K}{2R^2}\right)z' = 0, \qquad (4)$$

where the primes denote differentiation with respect to the scaled radius $R = r/u$, and

$$u = \frac{r_0}{2\sqrt{\Sigma}}, \quad \nu = \frac{K}{2\sqrt{\Sigma}} - \sqrt{1 + \frac{K}{2}} - \frac{1}{2}. \qquad (5)$$

The general solution of Eq. (4) reads $z'(R) = c_1 f_1(R) + c_2 f_2(R)$, where $c_1$ and $c_2$ are arbitrary constants and

$$\begin{aligned} f_1(R) &= e^{-R/2}R^{\sqrt{1+K/2}}U\left(-\nu, 1+\sqrt{4+2K}, R\right), \\ f_2(R) &= e^{-R/2}R^{\sqrt{1+K/2}}M\left(-\nu, 1+\sqrt{4+2K}, R\right) \end{aligned} \qquad (6)$$

The confluent hypergeometric functions of the second kind $U$ and $M$ are defined in Ref. [14]. Non-zero solutions of this form satisfying the boundary conditions only exist for certain values of the parameters, thereby defining the buckling threshold. Two parameter regimes must be distinguished:

- For $\nu \notin \mathbb{N}$, we have the following asymptotic behaviors:

$$\begin{aligned} f_1(R) &\underset{R \to 0}{\sim} \frac{\Gamma\left(\sqrt{4+2K}\right)}{\Gamma(-\nu)}R^{-\sqrt{1+K/2}}, \\ f_2(R) &\underset{R \to +\infty}{\sim} \frac{\Gamma\left(1+\sqrt{4+2K}\right)}{\Gamma(-\nu)}R^{-1-\nu-\sqrt{1+K/2}}e^{R/2} \end{aligned} \qquad (7)$$

Thus $f_1$ diverges as $R \to 0$ while Eq. (6) implies $f_2(0) = 0$. Hence the boundary condition $z'(0) = 0$ imposes $c_1 = 0$. Similarly, $f_2(R)$ diverges as $R \to +\infty$, thus $z'(+\infty) = 0$ yields $c_2 = 0$. Therefore there is no non-zero solution to the linearized buckling problem.

- For $\nu = n \in \mathbb{N}$, the singular terms of Eq. (7) vanish and $f_1$ and $f_2$ are both proportional to the generalized Laguerre polynomials $L_n^{(\alpha)}(R)$ [14]. Hence Eq. (4) has a unique solution, up to an arbitrary amplitude $\mathcal{C}$:

$$z'_n(R) = \mathcal{C}e^{-R/2}R^{\sqrt{1+K/2}}L_n^{\left(\sqrt{4+2K}\right)}(R). \qquad (8)$$

Since $L_n^{(\alpha)}(R)$ is a polynomial of degree $n$ in $R$, $z'_n$ satisfies the boundary conditions for any $n$. Therefore, there is an infinity of buckling thresholds, one per integer $\nu = n$. This is again reminiscent of the buckling rod problem, as each normal mode $z_n$ of the dressed membrane has its own instability threshold (Fig. 2). In the following we only consider the most unstable mode $n = 0$.

We now study strongly deformed dressed membranes [Fig. 1(c)]. We first check that our model accounts for the existence of long dressed membrane tubes similar to those observed in Refs. [7, 10, 11]. For a cylindrical protrusion of radius $r_t$ and length $\ell \gg r_t$, one can neglect the rounded tip and base of the tube. Introducing a fictitious vertical point force $f$ pulling the membrane up at $r = 0$, we minimize the free energy $\mathcal{G} = \mathcal{F} - f\ell$ with respect to $r_t$ and $\ell$ and find

$$r_t = r_0\sqrt{\frac{1+K}{2\Sigma}}, \quad f_t = \frac{2\pi\kappa}{r_0}\left[\sqrt{2(1+K)\Sigma} - K\right]. \qquad (9)$$

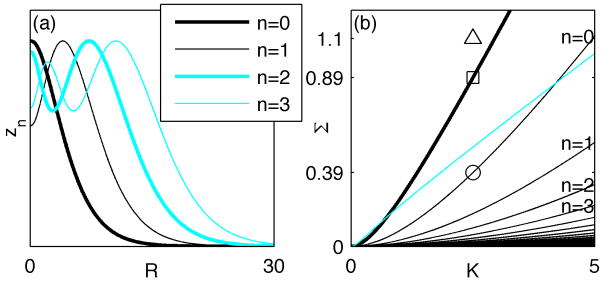

Figure 2: (color online) Normal modes $z_n(R)$ of the dressed membrane. (a) Spatial structure of the first four normal modes at their respective buckling thresholds for $K = 2.5$. (b) Thin black lines: buckling thresholds as a function of $n$, $K$ and $\Sigma$. Thick black line: stability limit of long, cylindrical dressed membrane tubes. Protrusions are obviously more stable at small $\Sigma$, where the destabilizing influence of the filaments overrides the stabilizing effect of the membrane. Therefore the $n$th normal mode of the flat dressed membrane is linearly unstable for parameter regimes located under the $n$th thin black line and long tubes exist only under the thick black line. Thin cyan (grey) line: parameter regimes compatible with the experimental data of Ref. [7]. Symbols are referred to in the main text.

Consider an equilibrium situation in which a long dressed membrane tube is held at a constant length by a force $f = f_t$. The force is then suddenly set to zero. In the case of an upward initial force $f_t > 0$, the tube tends to retract. If $f_t < 0$, on the contrary, $\ell$ increases and the dressed membrane spontaneously tubulates. This corresponds to the region of Fig. 2(b) located under the thick black line. Interestingly, long tubes are always stable when the flat dressed membrane is linearly unstable, but the reverse is not true. Thus there exists a regime, located between the thick black line and the $n = 0$ line of Fig. 2(b), where the flat dressed membrane is metastable. This regime is compatible with biologically reasonable parameter values. Indeed, combining Eqs. (3) and (9), one finds:

$$\Sigma = \frac{2(\kappa r_e r_t)^2 K^2 (1 + K)}{[\kappa r_e^2 + 2(\gamma - \sigma r_e) r_e r_t^2 + \kappa(r_e^2 + r_t^2) K]^2}. \quad (10)$$

Inserting $r_t \simeq 70$ nm and $r_e \simeq 200$ nm [7] and the estimates $\kappa = 20 k_B T$, $\sigma = 10^{-5}$ N.m$^{-1}$ and $\gamma = 1$ pN in this equation, we obtain a numerical relation between the scaled tension and filament stiffness characterizing the experiments of Ref. [7]. We plot this condition as a thin cyan (grey) line on Fig. 2(b). This line traverses both the metastable and unstable regions, making it possible that the experiments of Ref. [7] reflect either regime.

We consider the possibility that the flat dressed membranes observed in Ref. [7] are indeed metastable. In this hypothesis, an important quantity is the energy barrier $\Delta \mathcal{F}$ separating the flat state from the more stable, tubulated state. To compute $\Delta \mathcal{F}$, we numerically solve the full nonlinear shape equation of the tube on a finite domain $0 < R < 25$ [15]. Using $Z = z/u$, we define $S$ as the arc length along the dressed membrane in the $(R, Z)$ plane [Fig. 3(a)]. We parametrize the dressed membrane by $R(S)$ and the angle $\psi(S)$ defined by

$$\dot{R}(S) = \cos \psi(S), \quad \dot{Z}(S) = -\sin \psi(S), \quad (11)$$

where the dots denote the differentiation with respect to $S$. Minimization of the free energy $\mathcal{G}$ yields the shape equation of the dressed membrane:

$$\ddot{\psi} = \left( \frac{1}{4} - \frac{1 + \sqrt{4 + 2K} + 2\nu}{2R} + \frac{\cos^2 \psi + 1 + K}{2R^2} \right) \tan \psi$$
$$- \frac{\dot{\psi}^2 \tan \psi}{2} - \frac{\dot{\psi} \cos \psi}{R} - \frac{F}{R \cos \psi}, \quad (12)$$

where $F = fu/(2\pi\kappa)$. This equation is identical to Eq. (4) in the small-$\psi$ limit and to the bare membrane tube shape equation in the absence of protein ($k = 0$, $\mu = 0$) [15]. In the following, we discuss the specific example $K = 2.5$, but we believe that other values of $K$ yield a similar behavior. Let us first comment on the three regimes presented in Fig. 3(a-b). For $\Sigma = 1.1$ [indicated by $\triangle$ in Fig. 2(b)], tubes always retract in the absence of an external force, as shown in Fig. 3(b). Lowering the surface tension to $\Sigma = 0.89$ ($\square$), one reaches the boundary of the metastable region. For $0.39 < \Sigma < 0.89$, a positive force is required to extract short tubes, but long tubes grow spontaneously unless opposed by a negative $F$. At $\Sigma = 0.39$ ($\bigcirc$) and lower, even short tubes grow spontaneously and can be maintained at a finite length only by a negative force. In Fig. 3(b), crossings of the horizontal axis by the force-extension curves denote solutions of the biologically relevant, $F = 0$ problem, the stability of which is indicated by the sign of the curve's slope. Plotting the lengths of these protrusions as a function of $\Sigma$, we obtain the diagram Fig. 3(c), where we observe that the loss of stability of the $n = 0$ mode studied above yields a subcritical bifurcation. Focusing on the metastable regime ($0.39 < \Sigma < 0.89$), we note that forming an infinitely long tube requires first extruding a short tube from the dressed membrane, which is energetically unfavorable. The associated energy barrier $\Delta \mathcal{F}$ is given by the free energy of the unstable solutions represented by the main thin branch of Fig. 3(c). Integrating force-extension curves similar to those of Fig. 3(b), we calculate the work required to reach these solutions from the metastable, flat state and plot the results on Fig. 3(d). Under the effect of thermal fluctuations, an energy barrier of height $\Delta \mathcal{F}$ is crossed at a rate $\tau^{-1} e^{-\Delta \mathcal{F}/k_B T}$, where $\tau \sim$ ns is the characteristic relaxation time scale of the system. When $\Delta \mathcal{F}$ is of the order of a few $k_B T$, thermal fluctuations are sufficient to ensure the buckling of the dressed membrane within experimentally observable time scales. This is, however, not the case here, and the large energy barrier makes thermally activated ESCRT-III-mediated budding extremely unlikely in most of the metastable regime. Therefore, *in vivo*, ESCRT-III-mediated budding either takes place only in (or close to) the regime where the flat dressed membrane is linearly unstable, or is assisted by some unknown active process (*e.g.* actin polymerization, which is regulated by the ESCRT-associated protein Alix [16]).

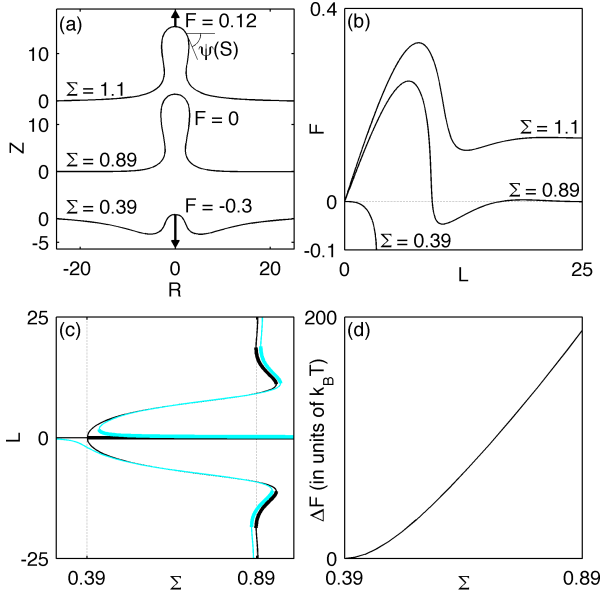

Figure 3: (color online) Numerically computed mechanical properties of strongly deformed dressed membranes for $K = 2.5$. (a) Parametrization and profiles. (b) Force-extension curves ($L = \ell/u$). (c) Black lines: bifurcation diagram for the $F = 0$ problem. Cyan (grey) lines: changes induced by a weak asymmetry of the dressed membrane. In both cases, thick (thin) lines represent stable (unstable) solutions. (d) Activation energy $\Delta\mathcal{F}$ a flat dressed membrane needs to reach the $\ell = +\infty$ buckled solution.

We now comment on two approximations used throughout this work. First, we assumed that the interactions between filaments and between filament and membrane are independent of the slope of the dressed membrane (*i.e.* of whether the filaments lie in the same plane or are stacked upon another). For small slopes, this dependence can be expanded as $\mu(\nabla z) = \mu_0 + \mu_2(\nabla z)^2/2 + \mathcal{O}\left[(\nabla z)^4\right]$ and yields the same linear stability analysis as above provided we redefine $\Sigma = (\sigma - \mu_0 - \mu_2)r_0^2/\kappa + k/(2\kappa)$. Second, we ignored in Eq. (1) any terms violating the $z \to -z$ spatial symmetry. These terms are allowed in general since the dressed membrane is not up-down symmetric, and might be responsible for the fact that buckling systematically occurs toward the outside of the cytoplasm [7]. Formally, such an asymmetry destroys the bifurcation studied here. If it is weak, however, a stable, almost flat configuration still exists for high tensions and loses stability close to the predicted $\Sigma = 0.39$ threshold, as illustrated in Fig. 3(c).

Finally, we believe that a better understanding of ESCRT-III-mediated budding could be gained by studying it experimentally in the absence of any active process. We propose an *in vitro* setup where an aspiration pipette is used to control the tension $\sigma$ of a giant unilamellar vesicle [17]. Introducing ESCRT-III proteins in the surrounding solution at a known concentration (and therefore at known $\mu$) [11], one could vary $\Sigma$ through $\sigma$ and directly measure the buckling threshold and its dependence on $\mu$. Due to the existence of the metastable region, we also predict a hysteretic behavior.

In this Letter we presented a novel buckling mechanism relevant for a wide range of systems involving interacting membranes and curved filaments (possibly including *e.g.* the one studied in Ref. [18]). Our robust qualitative and quantitative predictions can be tested in rather simple *in vitro* experiments and could shed light on the biological problem of ESCRT-III-mediated budding.

We thank Imre Derényi for help with the numerical procedures, Arnaud Échard for drawing our attention to ESCRT-III budding, Jacques Prost for constructive criticism and Aurélien Roux for support, discussions and experimental eagerness. We are also grateful to them and Markus Basan, Andrew Callan-Jones and Thomas Risler for critical reading of the manuscript.